\documentstyle[epsfig]{elsart}
\begin{document}
\begin{frontmatter}
\title{Fine Structure in the Energy Spectrum of Cosmic Ray Protons at
50 GeV ?}

\author[Moscow,Durham]{A.D.Erlykin}
\author[Kerman,Durham]{S.J.Fatemi}
\author[Durham]{A.W.Wolfendale}
\address[Moscow]{P.N.Lebedev Physical Institute, Leninsky pr. 53,
Moscow 117924, Russia}
\address[Kerman]{Physics Department, Shahid Bahonar University of
Kerman, Kerman, Iran}
\address[Durham]{Physics Department, University of Durham, Durham DH1
3LE, UK}

\begin{abstract}
The recently published precise spectrum of cosmic ray protons from the
Alpha Magnetic Spectrometer has been examined in some detail from the
standpoint of a search for deviations from a smooth, simple, power
law. We find a significant excess (~$\simeq$ 10\% over $\sim$ 0.3
interval in logE~) centered on 50 GeV in the published data. It is
possible that the 'unfolding technique' adopted by the experimenters
causes an overestimate of the excess but it is difficult to reduce it
much below about (~5$\pm$2~)\%.

\noindent We have examined other recent data, too. There is also
evidence here, for an excess in the same energy region, although of only
(~1.6$\pm$0.9~)\%. A value of (~3$\pm$1.5~)\% would be consistent with 
all the data. There are hints of similar excesses for heavier nuclei
and electrons. 

\noindent Possible explanations are put forward for an excess, should it prove
to be genuine.
\end{abstract}
\end{frontmatter}

\section{Introduction}

It is commonly asserted that the energy spectrum of cosmic rays is
almost featureless, being characterised by a power law spectrum which
slowly steepens near 3$\cdot$10$^{15}$ eV (~the knee~) and continues
with a new exponent to near 10$^{19}$ eV, where it flattens somewhat
(~the ankle~).

In fact, we (~Erlykin and Wolfendale, to be referred to as EW~) have
claimed that near the knee the situation is more complex
(~\cite{EW1} and references therein~). The complexity is attributed to
the effect of a single, local and recent supernova, the remnant of
which has accelerated particles, particularly oxygen and iron nuclei,
with a spectrum of the form E$^{-2}$ up to a rather sharp maximum at
$\simeq 4 \cdot 10^{14}Z$ eV. It is not unreasonable, statistically 
\cite{Fatem}, that such a single 'source' should show itself in a
rather narrow band of energy.
\section{The AMS experiment}
The EW result has led us to examine the situation at other energies,
viz. to search for irregularities. The very recent results from the
powerful AMS experiment \cite{Alcar} make a study of the proton
spectrum recorded by that experiment of considerable value, from the
standpoint of small deviations from smoothness. That such
irregularities cannot be more than a few tens of percent is inferred
from studies of the spectra reported by earlier workers (~to be
considered later~). However, the AMS (~satellite~) experiment, with
its 10$^7$ particles and carefully calibrated equipment, holds out the
hope of searching for irregularities down to a few percent in
intensity (~over some tens of GeV~).

The AMS authors have presented data on the proton spectrum from the
kinetic energy 0.1 to 200 GeV. Below about 10 GeV (~for certain
directions of incidence with respect to the earth~) interesting
geomagnetic effects are visible, but these are not the subject of
concern here. Rather, we examine the spectra above 10 GeV where
geomagnetic effects quickly become negligible. In this energy range
all the data presented,
which relate to 10 bands of geomagnetic latitude, are completely
consistent and, when averaged by us, they give the spectrum shown in
Figure 1.

\begin{figure}[htbp]
\begin{center}
\mbox{\epsfxsize=10cm\epsfysize=7cm\epsffile{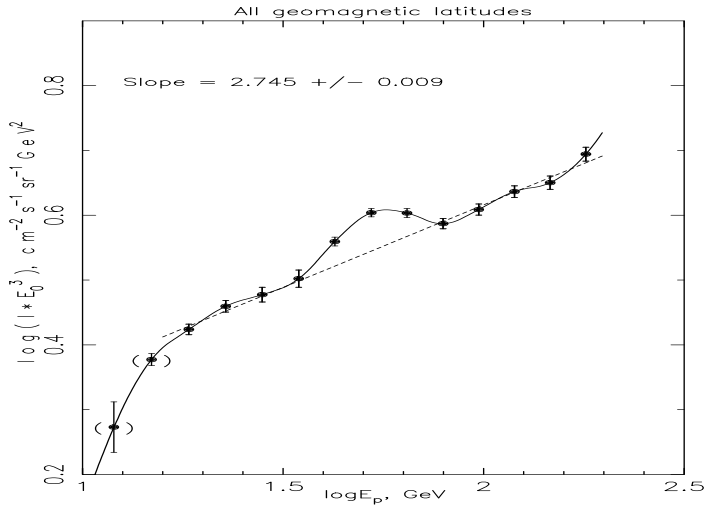}}
\end{center}
\caption{The AMS primary proton energy spectrum \cite{Alcar}
for all geomagnetic latitudes. The dashed line is the best fit of the
spectrum excluding the points in brackets, and those in the logE$_p$ = 
1.55 - 1.85 interval. The errors indicated are only statistical; those
in brackets are affected by geomagnetic processes. The magnitude of
the excess may be too high, by a factor $\sim$ 2 for reasons to do with the
corrections applied by the workers themselves (~see text~).}

\end{figure}

A discussion of the 'errors' in the intensities is necessary. In the
AMS work the errors for each of the ten bands of geomagnetic latitude
are enlarged considerably to allow for 'systematic errors'. Such
errors are, no doubt, present, but they should vary slowly from one energy
bin to the next - viz. they should not, in themselves, be able to give
the observed excess. The dispersion of the intensities (~the actual
points~) about their mean (~from the averaged set of ten~) is very
small indeed and the errors indicated by us in figure 1 are
correspondingly small. 

The result is of considerable interest in that it shows that a single
power law (~shown dashed~) will not fit the data. Instead, an excess
appears centered on 50 GeV. The 'peak' is some 15\% above the power
law of constant exponent neglecting the region of the peak.

It has been pointed out to us (~by Professor Yu.Galaktionov of the AMS
group~) that the unfolding technique used to 'correct' the data may
amplify existing irregularities. This is, of course, true but the fact
that every one of the ten spectra presented shows such (~amplified ?~)
peaks indicates, to us, that there is an underlying genuine peak. The
AMS group shows, as an example, a representative spectrum before and
after unfolding. We estimate that the 'raw' spectrum has an excess
about one half of that after correction. It seems safer, therefore, to
take this value as the AMS result, i.e. $\sim$ 6\% above the power law.

In Figure 2 we give the excess from a datum line which is drawn
through intensities in the range logE = 1.20 - 1.55 and 1.85
onwards, the intensities below logE = 1.20 are increasingly affected
by geomagnetic processes and are not used. An important point about
taking excesses
from such a datum is that the slowly varying systematic errors disappear on
subtraction. Thus, the only errors left in the AMS excesses of Figure
2 should be the, presumably smaller, rapidly varying systematic
errors. As remarked above, such errors due to the unfolding technique
probably reduce the AMS values by a factor 2, to give the line  
as shown dashed in the Figure.

\begin{figure}[htbp]
\begin{center}
\mbox{\epsfxsize=10cm\epsfysize=7cm\epsffile{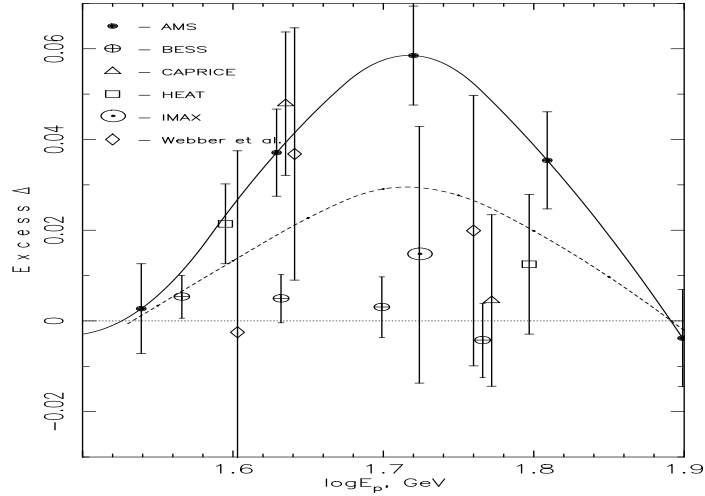}}
\end{center}
\caption{Excess of the primary proton intensities over the best fit
straight line: $\bullet$ - AMS \cite{Alcar}, $\bigoplus$ - BESS
\cite{Sanuki}, $\bigtriangleup$ - CAPRICE \cite{Boezio}, $\Box$ -
HEAT \cite{Swordy}, $\bigodot$ - IMAX \cite{Menn}, $\Diamond$ - Webber
et al. \cite{Webber}}
\end{figure}

Returning to the initial excess, the peak carries with it 
$\sim 2 \cdot 10^{-3}$ eV cm$^{-3}$, i.e. 0.4\% of the total cosmic
ray energy density; the reduced value is, correspondingly, 
$\simeq 1 \cdot 10^{-3}$ eV cm$^{-3}$.

An interesting feature is that the peak is rather narrow, being not far
from what would have been expected from the momentum resolution of the
instrument itself.
\section{Comparison with the Results of Other Workers}
\subsection{Results for Protons.}
It is necessary to examine the extent to which the AMS 'excess' is
also found in those other experiments in which the proton spectrum
has been measured. Clearly, despite the undoubted
superiority of the AMS project (~in terms of particle numbers, at
least~), it is necessary to have support from other work, or at least
reasons why others did not see the excess, before taking the excess
seriously.

Figure 2 shows the excess intensities derived by us from the data.
The logic in drawing a straight line for the datum needs
consideration, the main reason is that this is standard procedure as a
way of quantifying the spectral shape. There might, in fact, be some
very small curvature (~convex upwards~), but due to small residual solar
curvature and the narrow interval, $\sim$ 1.2/2.4, where the excess
was derived, the overestimate of the excess cannot exceed 0.003.
\subsection{The Results for Heavier Nuclei.}
The fraction of nuclei in the cosmic radiation falls with increasing Z
and the statistical accuracy falls accordingly. The search for fine
structure is, correspondingly, even more difficult than for
protons. Nevertheless, a search has been made.

A useful summary \cite{Shiba} has been given for the spactra of P, He,
C, O, Mg, Si and Fe in terms of rigidity. Such a parameter is
important in view of the fact that a promising explanation for the
proton structure is in terms of acceleration effects by way of weak
SNR in the ISM (~see \S 5~) and these will presumably be
rigidity-dependent. Inspection of a superposition of the rigidity
spectra over the range 100 - 200 GV reveals no peak at 50 GV (~
specifically $\Delta \leq$ 0.05, but perhaps a peak at twice this rigidity:
$\Delta = 0.07\pm0.05$ at 100GV.

Comparison with Figure 2 shows that the 50GV upper limit is a little above
the 'corrected' AMS value (~$\Delta \simeq$ 0.03~) and thus not
inconsistent with it. However, it gives no positive support to the
proton results. 

Turning to 100 GV rigidity (~100 GeV energy for protons~) inspection
of Figure 1 shows a peak there of less than about 0.01, not
inconsistent with the 0.07$\pm$0.05 for the heavy nuclei, but again
not providing positive support. It should be remarked, however, that
100 GeV is getting close to the 'maximum detectable energy' of the AMS
spectrometer, and peaks would certainly be smeared here and their
magnitude considerably reduced.
\subsection{The Results for Electrons.}
It is here that recent work \cite{Torii} gives rather strong evidence
for a peak at 50 GeV, of magnitude $\Delta = 0.27 \pm 0.1$. The only
other experiments of adequate statistical precision give a broad peak
in the same region of smaller magnitude: in one case $0.1\pm0.1$
\cite{Tang} and a null result in the other: $0\pm0.1$ \cite{Silver},
the last-mentioned being a measurement reported 24 years ago. The
average is $\Delta = 0.10\pm0.06$. 
\section{Discussion of the results}
Taking the AMS results at their face value the 15\% excess in the peak
reduces to (11$\pm$1.5)\% when averaged over the range logE: 1.55 -
1.85. However, as remarked earlier, a safer estimate is lower by a
factor $\sim$ 2, with a larger error; specifically the 'best estimate'
for the average is (5 $\pm$ 2)\%.

Turning to the mean of the other results, this is (~1.6$\pm$0.9)\%, a
value lower than our AMS value but not impossibly so. A figure of 
(3$\pm$1.5)\% would be consistent with both. 

Turning to the results for other components, and transforming from
$\Delta$-values to percentages, we have:
\begin{itemize}
\item nuclei of the same rigidity: $<$ 12\%
\item electrons of the same rigidity ( and energy ): $26\pm16$\%
\end{itemize}
The results for nuclei are not inconsistent with the proton results
and the electron 'peak' is, if anything, higher. Taken with the 100 GV
nuclear excess (~of $17\pm12$\%~) there is at least the suggestion of
small peaks in the other components, too.
\section{Possible explanations}
A non-exhaustive list is as follows.\\
(i) {\em Technical origin}\\
It is unlikely that technical factors, such as errors in exposure at
the energy in question, are responsible for the excess in {\em all}
the measurements, but technical causes cannot be ruled out.\\
(ii) {\em Protons from exotic processes}\\
Concerning the AMS excess the easiest (~observationally~) explanation
would be a delta-function, i.e. the presence of protons of unique
energy near 50 GeV. Apart from the large energy density (~0.2-0.4\% of
the whole cosmic ray energy density~) the absence of anti-protons of
the same flux (~at least none have been reported~) acts against some
exotic processes, such as the decay of massive baryons and
anti-baryons, at least.\\
Other possibilities involve the annihilation of dark matter particles
of supersymmetric type \cite{Kamio}. Again, the non-observation of
accompanying particles - such as gamma rays - rules out many
possibilities. Nevertheless, there {\em may} be appropriate decay
modes which would be acceptable. This aspect is, in fact, topical
because of the possibility of dark matter candidates having the
appropriate mass.\\
(iii) {\em Supernova Remnant Effects}\\
Although not predicted by acceleration models, such as was - in a
sense - the situation for the EW mechanism at PeV energies, this is
perhaps the best contender. The stochastic nature of supernova shocks
means that there {\em must} be spectral structure at some, small,
level, at all energies where supernovae play a part. We are currently
examining this aspect in some detail.\\
The likely presence of small effects for the spectra of nuclei and
electrons increases the attractiveness of this interpretation.\\
(iv) {\em Heliospheric Shock Effects}\\
The energy in question, 50 GeV, is of the right order to be possibly
accomodated by shocks at the heliospheric boundary. However, why the
excess should be so sharp (~after allowance for resolution
broadening~) is a mystery. 
\section{Conclusions}
Inspection of the very recent, precise, AMS measurements of the
primary cosmic ray proton spectrum shows an interesting excess in the
region of 50 GeV. Although the quoted values may overestimate the
excess - perhaps by a factor 2 - it seems unlikely that the whole
effect will go away.

Including results from other proton experiments, too, an overall excess
averaged over $\Delta$ logE = 0.15 centred on 50 GeV of
(~3$\pm$1.5)\% would fit all the data.

It should be possible to confirm or deny a value of this magnitude by
a more sophisticated study of the AMS results.

More importantly, perhaps, is the general point that modern experiments
of apparently high precision, should focus on examinations of the
detailed shape of the energy spectrum of protons, of heavier nuclei
and of electrons, too. Fine structure should be visible at some level;
perhaps it is already starting to be seen ?\\
Concerning the explanation, if the heavier nuclei do, in fact, show
similar effects to those for protons (~at present the results are
clearly very marginal~) - at the same rigidity - it will be possible
to rule out exotic processes as being responsible. Weak, local
supernova remnants, would then be the preferred mechanism. However, it
is premature to rule out exotic processes, yet.\\
{\bf Acknowledgements}\\
We are grateful to Professor Yu.Galaktionov for useful discussions and
comments. The Royal Society is thanked for the provision of financial
support by way of a Joint Project. Mr. P.Kiraly kindly provided
helpful comments, as did the referee, and we are grateful to them.

\end{document}